\documentclass[conference]{IEEEtran}
\usepackage{cite}
\usepackage{amsmath,amssymb,amsfonts}
\usepackage{algorithmic}
\usepackage{graphicx}
\usepackage{textcomp}
\usepackage{xcolor}
\usepackage{fancyhdr}
\usepackage[hyphens]{url}

\usepackage{tikz}
\usepackage{xcolor}
\usepackage{subfigure}
\usepackage{color}
\usepackage{pifont}

\newcommand{\tabincell}[2]{\begin{tabular}{@{}#1@{}}#2\end{tabular}}
\newcommand*{\circled}[1]{\lower.7ex\hbox{\tikz\draw (0pt, 0pt)%
		circle (.5em) node {\makebox[1em][c]{\small #1}};}}

\def\BibTeX{{\rm B\kern-.05em{\sc i\kern-.025em b}\kern-.08em
    T\kern-.1667em\lower.7ex\hbox{E}\kern-.125emX}}

\fancypagestyle{firstpage}{
  \fancyhf{}
 
  \fancyfoot[C]{\thepage}
}

\title{Update the Root of Integrity Tree in Secure Non-Volatile Memory Systems with Low Overhead} 
\author{\IEEEauthorblockN{Jianming Huang and Yu Hua}
	\IEEEauthorblockA{
		\textit{Huazhong University of Science and Technology}\\
		\textit{Email: \{jmhuang, csyhua\}@hust.edu.cn} }
}

\begin{document}
\maketitle
\thispagestyle{firstpage}
\pagestyle{plain}


\begin{abstract}

Data integrity is important for non-volatile memory (NVM) systems that maintain data even without power. The data integrity in NVM is possibly compromised by integrity attacks, which can be defended against by integrity verification via integrity trees. After NVM system failures and reboots, the integrity tree root is responsible for providing a trusted execution environment. However, the root often becomes a performance bottleneck, since updating the root requires high latency on the write critical path to propagate the modifications from leaf nodes to the root. The root and leaf nodes have to ensure the crash consistency between each other to avoid any update failures that potentially result in misreporting the attacks after system reboots. In this paper, we propose an efficient and low-latency scheme, called SCUE, to directly update the root on the SGX integrity tree (SIT) by overlooking the updates upon the intermediate tree nodes. The idea behind SCUE explores and exploits the observation that only the persistent leaf nodes and root are useful to ensure the integrity after system failures and reboots, due to the loss of the cached intermediate tree nodes. To achieve the crash consistency between root and leaf nodes, we accurately predict the updates upon the root and pre-update the root before the leaf nodes are modified. Moreover, the SIT root is difficult to be reconstructed from the leaf nodes since updating one tree node needs its parent node as input. We use a counter-summing approach to reconstructing the SIT from leaf nodes. Our evaluation results show that compared with the state-of-the-art integrity tree update schemes, our SCUE scheme delivers high performance while ensuring the system integrity.

\end{abstract}

\vspace{-0.1cm}
\section{Introduction}
\label{section 1}

Non-Volatile Memory (NVM) has demonstrated the salient features of non-volatility, low power consumption and high performance. To guarantee the data security and confidentiality, we need to encrypt the data in NVM. Moreover, the data integrity is also important, which is interpreted that the data can't be illegally tampered with. The encrypted data need to be further verified to ensure the integrity, which requires security metadata, i.e., counter blocks in counter mode encryption and tree nodes in integrity tree verification. However, due to the persistence of NVM, these security metadata need to be crash consistent and recoverable, to guarantee the NVM systems continue to run safely and efficiently after system failures and reboots~\cite{ZubairA19}.

To ensure data confidentiality, existing works use the counter mode encryption (CME)~\cite{YanEPRS06} to encrypt the data in NVM. CME uses counters to generate the one-time padding (OTP) and XORs the OTP with the plaintext/encrypted data to generate the encrypted/plaintext data. CME fetches the counter blocks in the metadata cache in the memory controller in advance. The generation of OTP is in parallel with reading the data from NVM. Therefore, the latency of decryption is hidden by that of reading data. The data integrity is verified by the integrity trees, such as merkle tree (MT), bonsai merkle tree (BMT)~\cite{RogersCPS07} and SGX integrity tree (SIT)~\cite{CostanD16}. In the MT, the user data are iteratively hashed to generate the Keyed-Hash Message Authentication Codes (HMACs) in the intermediate tree nodes and the root. When the user data are modified, the root of MT changes to reflect the modifications of the user data. The data read from NVM also needs to be verified by the cached intermediate tree nodes and the root.

However, MT fails to efficiently support the encryption in NVM. To combine the counter mode encryption with integrity verification, the bonsai merkle tree (BMT) is used. BMT treats the counter blocks as leaf nodes, and the intermediate nodes and root are generated by iteratively hashing the counter blocks. The counters increase to encrypt the new data when being persisted. Moreover, the modifications of the counter blocks in BMT are further propagated to the root. BMT elaborately combines the counter mode encryption and integrity tree by organizing all security metadata (i.e., counter blocks and integrity tree nodes) in the integrity tree. SIT also organizes the counter blocks as leaf nodes in the integrity tree. Unlike BMT, the intermediate tree node in SIT contains eight counters and one HMAC. To exhibit the changes of leaf nodes, the root in SIT needs to be updated in time while the modified leaf nodes are persisted.

The root in the integrity tree is the only trusted base to verify the integrity of data. Since the root is stored in the non-volatile register on chip, the root can't be attacked. Other tree nodes will be flushed into NVM and tampered by the attackers, and thus the tree nodes in NVM are not trusted. When running the system, the cached intermediate tree nodes can be regarded as the trusted bases since they are directly/indirectly verified by the root. However, when the system failures occur, the cached intermediate nodes are lost, and only the root is trusted. For verifying the leaf nodes after system reboots, we need to timely update the root during the running time. 

However, updating root in the integrity tree needs to address two problems: \circled{1}\textbf{the long root update latency:} the long latency of propagating the modifications from leaf nodes to root on the critical path, and \circled{2}\textbf{the root crash inconsistency:} the crash inconsistency between root and leaf nodes when the system failures occur. Moreover, \textbf{SIT can't be reconstructed from leaf nodes} since constructing one SIT node requires the counter in the parent node as input. 

When the leaf nodes are modified, the modifications need to be propagated to the root for the verification after system failures and reboots. However, propagating the modifications in MT/BMT, i.e., iteratively hashing the data, incurs a long latency. One hash computation needs 80--160 cycles~\cite{liu2019janus,suh2003efficient,gassend2003caches}. Due to the large capacity of NVM, the height of the integrity tree is high, e.g., tens of levels in the integrity tree for the 16GB NVM. Propagating the modifications needs tens of hash computations incurring a long latency of the systems. SIT has the ability to compute the HMACs in parallel. However, SIT needs multiple hash circuits to compute multiple HMACs while the integrity verification is performed in the memory controller with limited computational resources~\cite{Zuo0ZZG18}, which consumes the expensive on-chip space and incurs energy consumption. Since NVM can provide TB-scale capacity and the integrity trees in the large NVM are much high~\cite{kadekodi2019splitfs}, it is important to decouple the latency of updating SIT root from the number of hash circuits.

Existing schemes overlook the system performance slowdown incurred by the root update latency of integrity tree. In secure NVM systems~\cite{ZubairA19,YangLCMS19,AwadYSNZ19}, the writes to user data are considered to be completed if the write requests arrive at the write queue, which is persistent domain due to the support of Intel Asynchronous DRAM Refresh (ADR) technique~\cite{ADR}. Propagating the modifications to update the integrity tree root is executed by the backend threads. Therefore, the update latency doesn't impact the system performance. However, due to the recoverability of the NVM systems, the root of the integrity tree must be updated in time before system failures to ensure the data integrity verification after system reboots. The write requests of user data are actually completed when the modifications propagate to the integrity tree root. The operation of updating root exists on the critical path and significantly reduces the system performance~\cite{liu2019janus}.

Like the counter and user data described in SCA~\cite{LiuKRK18}, there exist crash inconsistency problems between the root and leaf nodes. System failures may occur at any time. If the failures occur after persisting leaf nodes but before updating the root, the old root can't verify the new leaf nodes after system failures, which possibly causes the misreport of attacks. Similarly, if the failures occur after updating root but before persisting leaf nodes, the attacks are also misreported after system reboots. We call it \textit{root crash inconsistency} problem, i.e., after system failures, the root needs to be consistent with the leaf nodes.

After system failures, we use the root as the trusted base to verify other data. However, since the SIT can't be recovered from the leaf nodes, the root of SIT can't play its role after system reboots. The generation of one SIT node needs its parent node as input. Due to the loss of the intermediate nodes during failures, the nodes in the SIT can't be reconstructed from leaf nodes~\cite{ZubairA19}. After reboots, the system doesn't know whether the leaf nodes are attacked or not via verifying the SIT root. This drawback limits the usage of SIT. In existing SIT-based NVM systems~\cite{ZubairA19,alwadi2019phoenix}, one MT constructed in cache or NVM is added outside the SIT, and the root of the MT is used as the trusted base to verify the integrity of data. Some of other integrity-tree works~\cite{freij2020streamlining,AwadYSNZ19} can't be leveraged by the SIT since SIT fails to be reconstructed from leaf nodes.

We observe that in SIT, increasing the child counter causes the increment of the parent counter, and thus the parent counter is the sum of the corresponding child counters. Moreover, the counter in root is the sum of all corresponding leaf counters. Based on the observation, to reduce the long latency of updating root, we propose the \textbf{S}hort\textbf{C}ut \textbf{U}pdat\textbf{E} scheme (SCUE) to correctly update the root of SIT with low overheads, and leverage a counter-summing recovery approach to reconstructing the SIT from the leaf nodes. When propagating the modifications of leaf nodes, we directly update the root of SIT by overlooking the intermediate nodes, called SCUE scheme. In SCUE, the leaf nodes and root are guaranteed to be consistent, excluding the intermediate nodes. After system reboots, we reconstruct the whole SIT by the same observation that the parent counter is the sum of the corresponding child counters. The reconstructed root detects the attacks on the leaf nodes by comparing with the stored one. 

SCUE decreases the latency of updating root, and however, the failures also occur between persisting leaf nodes and updating the root. To ensure the root crash consistency, we observe that the root of SIT can be predicted. Thus we pre-update the root in SCUE to prepare the root in the ADR region.

To evaluate the performance of our proposed scheme, we use Gem5~\cite{BinkertBBRSBHHKSSSSVHW11} with NVMain~\cite{PorembaZ015} to implement SCUE. We evaluate 5 persistent workloads that have been widely used in the community~\cite{CoburnCAGGJS11,RenZKCWM15,KolliRDSPLCW16,KolliGSDCNW17,LiuKRK18,zuo2019supermem} and 8 macro-benchmarks from the SPEC2006~\cite{Henning06}. Our experimental results show that SCUE significantly reduces the write latency by 1.81x on average and provides 1.59x system speed up (up to 2.28x) compared with existing tree update schemes. 

In summary, this paper makes the following contributions:
\begin{itemize}
	\setlength{\itemsep}{0pt}
	\setlength{\parsep}{0pt}
	\setlength{\parskip}{0pt}
	\item \textbf{Shortcut update scheme for reducing write latency.} We propose a new update scheme to directly update the root of the integrity tree without propagating the modifications on the intermediate tree nodes.
	\item \textbf{Ensuring the root crash consistency.} We analyze the root crash inconsistency problem and pre-update the root in the ADR to address the problem.
	\item \textbf{Reconstructing the SIT from the leaf nodes.} We use counter-summing approach to recovering the SIT from the consistent leaf nodes without the aid of the MT. This work compensates for the shortcomings of SIT compared with the BMT.
	\item \textbf{Extensive experiments.} We have implemented and evaluated SCUE via micro- and macro-benchmarks. The experimental results show that our proposed scheme delivers high performance while ensuring the system integrity.
	
\end{itemize}

\vspace{-0.2cm}
\section{Background and Motivation}

\begin{figure}[t]
	\vspace{-0.3cm}
	\centering
	\includegraphics[width=0.45\textwidth]{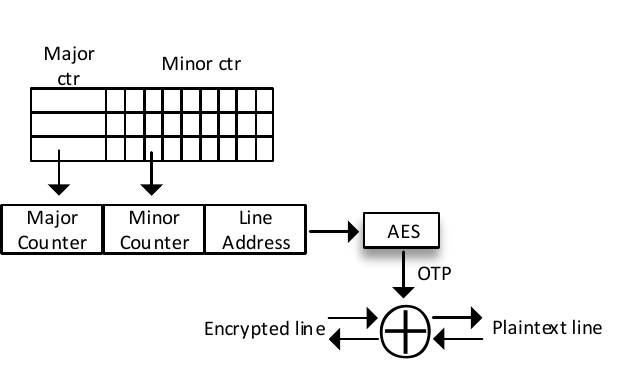}
	\vspace{-0.5cm}
	\caption{The Counter Mode Encryption scheme.}
	\label{CME}
	\vspace{-0.5cm}
\end{figure}

\subsection{Threat Model}
Existing works~\cite{YanEPRS06,RogersCPS07,AwadMHSH16,LiuKRK18,YeHA18,ZubairA19,AwadYSNZ19} assume that only the on-chip domain in the computer system is safe, including the processor, cache and memory controller, which we follow in our threat model. The NVM can be attacked to reveal the data, such as stolen DIMM and bus snooping attacks. The data confidentiality attacks can be defended by encryption~\cite{chhabra2011nvmm}. The memory tampering attacks modify the data in NVM to compromise the data integrity, including data replay attacks~\cite{sung1997shared}. These integrity attacks can be detected by the integrity trees. Other types of attacks are beyond the scope of this paper. In this paper, we mainly focus on the integrity trees to improve the performance of SIT.

\subsection{Counter Mode Encryption}
The counter mode encryption (CME) has been widely used in state-of-the-art security systems~\cite{AwadMHSH16,YoungNQ15,SwamiRM16,Zuo0ZZG18}. In general, to encrypt the data, direct AES encryption is used, which however places the decryption latency on the read critical path. Moreover, due to the unchanged secret key, the AES can be broken via the dictionary attacks. To deliver high performance and improve the security of the systems, the counter mode encryption is used as shown in Fig.~\ref{CME}. CME uses the counter and line address to generate the one-time-padding (OTP). When writing data, the ciphertext is generated by XORing the plaintext and OTP. When reading data, since the counter blocks are cached in the memory controller, systems generate the OTP and read the encrypted data in parallel. The encrypted data is decrypted by XORing the ciphertext and OTP. Therefore, the decryption latency is masked by the data read latency. 

For security considerations, the OTP should never be reused. CME uses the line address as OTP generation input to ensure that different data lines have different OTPs. When one modified data is persisted into NVM, the corresponding minor counter increases by 1. For the same data, the OTP is not reused in each write since the counters are different. One counter block contains one 64-bit major counter and 64 7-bit minor counters. When the minor counter overflows, the major counter increases by one, all minor counters in the counter block are reset to 0, and 64 corresponding user data blocks need to be re-encrypted.

\subsection{Integrity Verification}
\label{integrity verification}
Data integrity verification is important for system security. An attacker tampers with the user data without authorization, which becomes even worse once the tampered data are used by CPU. Normally, systems use Keyed-Hash Message Authentication Codes (HMACs) to verify the integrity of data. HMACs are generated by hashing the data, address and secret keys. When reading data, systems verify the data integrity via comparing the stored and recomputed HMACs. If the two HMACs are different, systems detect the unauthorized modifications. Due to the lack of secret keys, the attackers fail to construct the matched HMACs of the modified data and can't pass the integrity checking. 

However, HAMCs can't detect the data replay attacks that break the data integrity. Attackers record the old data and HMAC, and use the old tuple to replace the new data and HMAC. By only using HMAC, systems can't detect replay attacks, since the old HMAC matches the old data. To defend against the replay attacks, an integrity tree is used, including the Merkle Tree (MT), Bonsai Merkle Tree (BMT) and SGX Integrity Tree (SIT)~\cite{GassendSCDD03,RogersCPS07,CostanD16,TaassoriSB18}, to protect data via the on-chip root that can't be tampered by attackers.

\begin{figure}[t]
	\vspace{-0.3cm}
	\centering
	\includegraphics[width=0.45\textwidth]{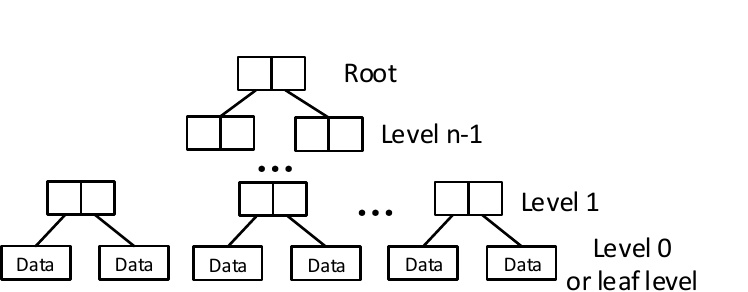}
	\vspace{-0.2cm}
	\caption{The merkle tree constructed from user data.}
	\label{MT}
	\vspace{-0.5cm}
\end{figure}

\begin{figure}[t]
	\vspace{-0.3cm}
	\centering
	\includegraphics[width=0.45\textwidth]{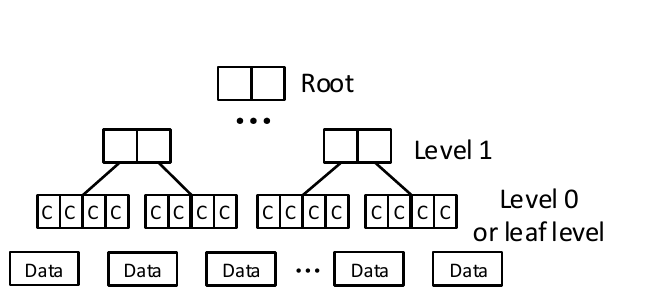}
	\vspace{-0.1cm}
	\caption{The bonsai merkle tree constructed from counter blocks.}
	\label{BMT}
	\vspace{-0.5cm}
\end{figure}

\vspace{-0.1cm}
\subsection{Merkle Tree}
Merkle trees (MT) are used to protect the data integrity by constructing the whole tree from the user data which are leaf nodes in the MT. In an 8-ary MT, 8 data are hashed to generate the upper-level node that consists of 8 HMACs, called \textit{parent node} of the hashed data. The upper-level nodes are also hashed to generate the higher-level nodes, and finally, the root is generated by iteratively hashing the leaf nodes. As shown in Fig.~\ref{MT}, the leaf nodes are in Level 0, i.e., leaf level, and they are iteratively hashed to construct the Level 1 and higher-level nodes. Systems store the root on the chip so that the root can't be tampered by attackers in our threat model.

Modifying one user data causes the change of its parent node in the MT, and the modification will propagate to the root by iteratively modifying the intermediate nodes. If attackers aim to replay the user data, they need to replay the parent node. Otherwise, the replay attacks are detected by the parent node. However, replaying the parent nodes can be detected by Level 2 parent nodes, and so on. Finally, achieving replay attacks needs to modify the on-chip root which is invulnerable to attackers. Thus the replay attacks can't succeed in the MT-based memory systems.

In NVM, due to the large capacity for storing user data, the number of leaf nodes in MT is large, causing a high MT. The overheads of storing tree nodes and propagating modifications from leaf nodes to root are expensive. To reduce the overheads, MT is combined with the CME in the security NVM systems. The counter blocks in CME are leaf nodes in the integrity tree, and iteratively hashed to generate the root as shown in Fig.~\ref{BMT}. This integrity tree is called Bonsai Merkle Tree (BMT)~\cite{RogersCPS07}. Since one counter block covers 64 data blocks, the number of the leaf nodes and the height of BMT are much smaller than that of MT. In BMT, if attackers replay data, they need to replay the counters since the data are connected with counters via the HMACs. Therefore, the replay attacks are detected by BMT, like the attacks in MT. Since the MT/BMT is constructed via the leaf nodes i.e., user data in MT and counter blocks in BMT, after system failures, we can rebuild the MT/BMT from the up-to-date and consistent leaf nodes.

\begin{figure}[t]
	\vspace{-0.3cm}
	\centering
	\includegraphics[width=0.45\textwidth]{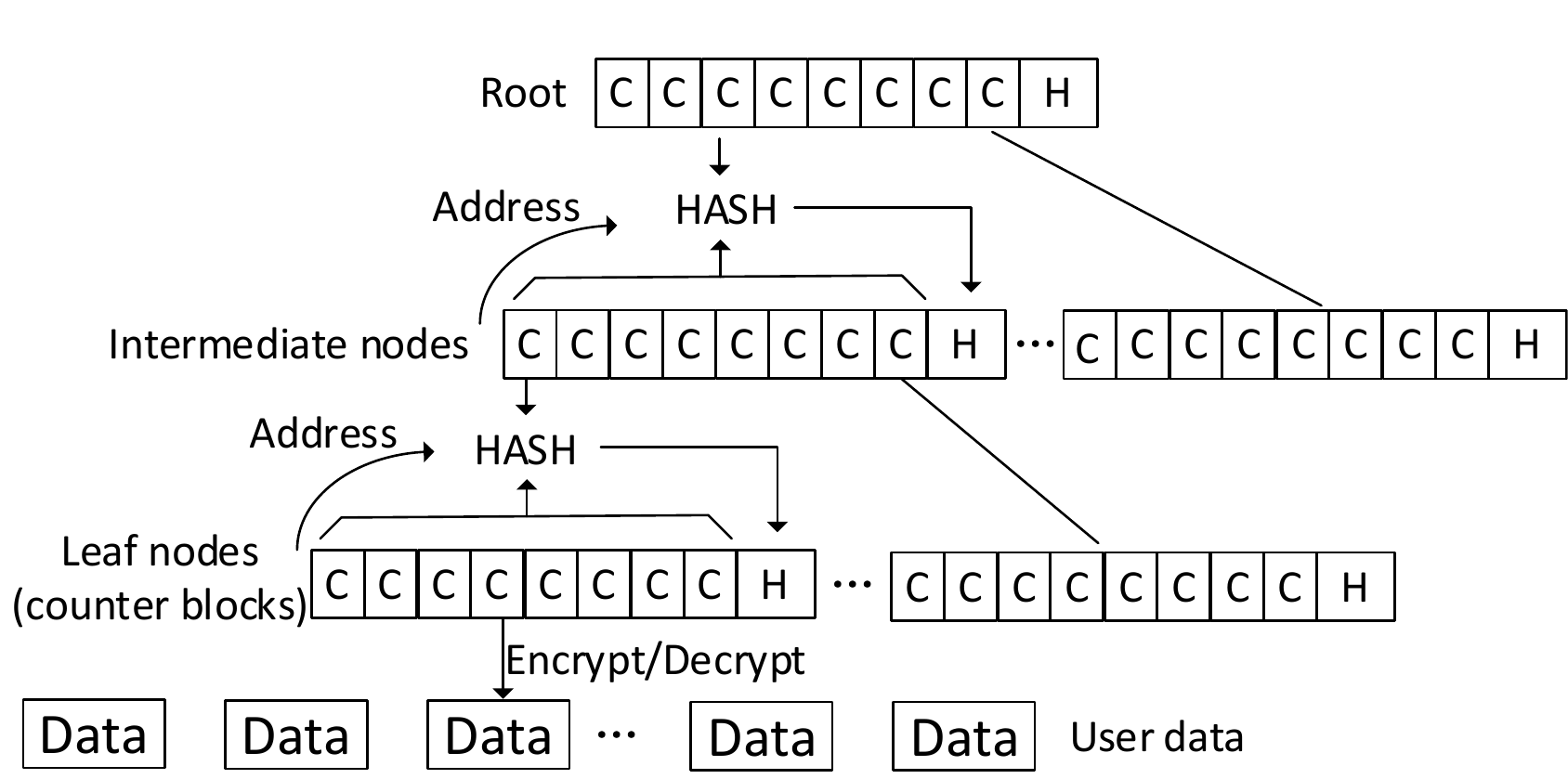}
	\vspace{-0.1cm}
	\caption{The SGX integrity tree.}
	\label{SIT}
	\vspace{-0.5cm}
\end{figure}

\vspace{-0.2cm}
\subsection{SGX Integrity Tree}  
Integrity verification schemes~\cite{CostanD16,TaassoriSB18,SaileshwarNREJQ18} often use SGX Integrity Trees (SIT). Unlike the nodes consisting of HMACs in the MT and BMT, the tree nodes in SIT consist of eight counters and one HMAC. In SIT, one tree node has eight child nodes corresponding to the counters in the node one by one. As shown in Fig.~\ref{SIT}, we compute the HMAC in the node by hashing the address of one node, all counters in this node, and one corresponding counter in the parent node. The node update way in SIT is similar to the counter mode encryption, i.e., persisting one modified node causes the corresponding counter in its parent node to increase by one, and the HMAC is also recomputed. 

For updating root, like BMT, SIT has two update schemes: eager and lazy schemes~\cite{alwadi2019phoenix}. The eager scheme immediately propagates the modifications of the leaf nodes to the root once the modified leaf nodes are persisted. In an eager scheme, persisting one modified leaf node incurs the changes on all nodes in the branch, including the root. The corresponding counters in these nodes increase by one. In a lazy scheme, the root is not immediately updated. The node in SIT is only modified when its modified child nodes are persisted. 

Both eager and lazy schemes can ensure the integrity of user data during running time. Since the cache is the secure domain, the cached nodes are treated as trusted bases like the root to verify other data. To implement the transmission of trusted base from root to the cached nodes, each data entering cache must be verified. When reading data, systems need to read the ancestor nodes until one ancestor node has already existed in cache. These ancestor nodes are iteratively verified to ensure the security of the read data.

SIT uses counters and HMACs to protect data. When the replayed user data and HMAC are read, systems recompute the HMAC by hashing the address, the data and the parent counter in counter block to verify the integrity of data. Attackers can also replay the counter block when it is persisted in NVM. However, the parent node of the counter block is updated in cache. Before systems use the replayed counter block to verify the attacked user data, the replayed counter block needs to be verified by the updated parent node, since the counter block re-enters the cache. In the worst case, attackers tamper with all nodes in a branch, if they are not in cache, to execute the replay attacks. However, since the root always resides in the chip, the updated root can be used to iteratively verify the integrity of the user data.

\subsection{Counter Crash Consistency}
Counter blocks need to be consistent with the user data. If the system failures occur between persisting user data and corresponding counter blocks, the unmatched counter blocks can't correctly decrypt the encrypted user data.

Osiris~\cite{YeHA18} doesn't force to persist the counter blocks with user data. The stale counter blocks can be restored by increasing the counter value. The correctness of the alternative counter values needs to be checked by recomputing and comparing the Error Correction Code (ECC) with the stored one. If the two ECCs are matched, the alternative counter is correctly recovered.

To ensure the counter crash consistency, which is beyond the scope of this paper, we use the Osiris in our work. In fact, other counter crash consistency schemes, such as Supermem~\cite{zuo2019supermem} and SCA~\cite{LiuKRK18}, can be also used.

\subsection{The Problem of Updating Root}
After system failures and reboots, the data integrity needs to be verified. Since the cached metadata are lost, the root is used as the trusted base. In general, systems reconstruct the tree from the leaf nodes. If the reconstructed root is the same as the stored one, the leaf nodes are not tampered by attackers. And the user data are protected by the leaf nodes.

To ensure the system integrity after failures, the root of the integrity tree needs to be timely updated. However, existing works~\cite{zuo2019supermem,LiuKRK18,YangLCMS19} don't discuss the challenges of updating the root. 

We observe two challenges when updating the root in the integrity tree. \circled{1}\textbf{The high overhead of updating root in an integrity tree.} To correctly verify the leaf nodes after system failures, the modifications of leaf nodes need to be propagated into the root before system failures. However, propagating the modifications incurs long latency on the write critical path. To update the root, all nodes in the branch are read from NVM (if they are not in cache), which consumes some latency, called \textit{read\_latency}. Since updating one node uses its child nodes as inputs, updating the root consumes a long latency to serially hash all nodes in the branch, called \textit{hash\_latency}. Hashing the inputs to generate one HMAC needs 80--160 cycles~\cite{liu2019janus,suh2003efficient,gassend2003caches}. To propagate the modifications from leaf nodes to root, we serially hash tens of HMACs, which incurs a long latency.

\begin{figure}[t]
	\vspace{-0.3cm}
	\centering
	\includegraphics[width=0.30\textwidth]{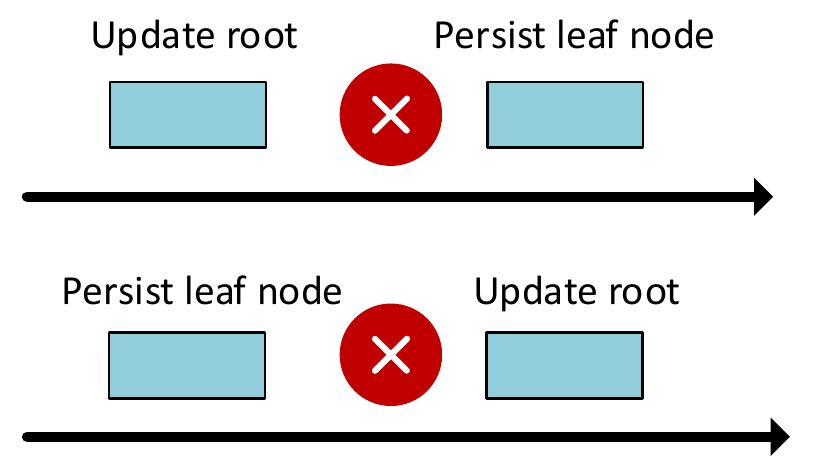}
	\vspace{-0.1cm}
	\caption{The failures occurring between updating root and persisting leaf node cause the inconsistency problem between root and leaf nodes.}
	\label{root-crash}
	\vspace{-0.3cm}
\end{figure}

~\circled{2}\textbf{The inconsistency between root and leaf nodes.} Since the write queue in the memory controller is supported by ADR, the write queue is a persistent domain. Updating the on-chip root and persisting the leaf nodes into the write queue are two atomic operations. As shown in Fig.~\ref{root-crash}, if a system failure occurs after updating root but before persisting leaf node, the new root can't correctly verify the old leaf nodes. On the other hand, if a system failure occurs after persisting leaf node but before updating root, the new leaf node also can't match with the un-updated root, thus causing \textit{root crash inconsistency problem}. When updating root and persisting a leaf node are executed in an atomic manner, the leaf nodes can be verified by the root after the system failures. After verifying the leaf nodes, i.e., counter blocks in BMT and SIT, the user data can be further verified via the HMAC.

SIT is a parallelizable integrity tree to significantly reduce the propagating latency. The HMACs in SIT nodes are computed in parallel once the counters are ready. However, parallel generating the HMACs in nodes requires multiple hash circuits, and consumes computing resources. Security schemes, including encryption and integrity verification, are executed in the memory controller with limited resources. Moreover, the capacity of NVM can be up to 6TB~\cite{nguyen2018picl}, which means tens of nodes exist in a branch. To generate HMACs in all nodes in a branch, tens of hash circuits in the memory controller become necessary, while incurring the consumption of on-chip space and energy. To decouple the root update latency from the number of hash circuits, it is important to consume low computing resources and reduce the latency of updating root in SIT.

Moreover, even if the root in SIT is correctly updated in time, the root can't verify the data integrity after system reboots. The root fails to be reconstructed from the leaf nodes~\cite{ZubairA19} to be compared with the stored one to detect the leaf nodes. Without the correct counters in the target node and its parent node, the target node can't be reconstructed. To make the SIT root play its role after system reboots, it is important to recover the SIT from persistent leaf nodes.

\begin{figure}[t]
	\vspace{-0.3cm}
	\centering
	\includegraphics[width=0.45\textwidth]{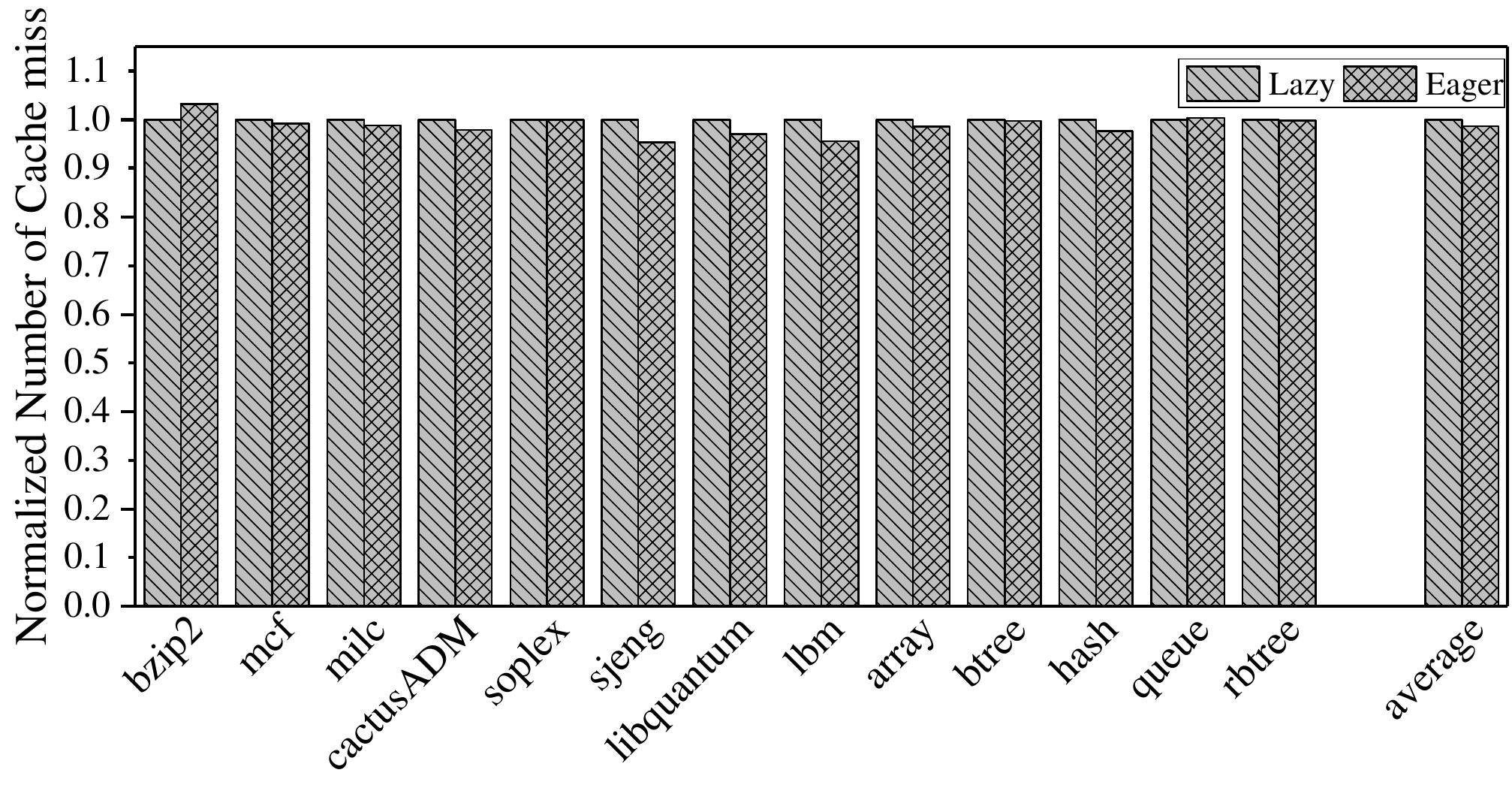}
	\vspace{-0.3cm}
	\caption{The numbers of metadata cache misses on lazy and eager schemes in SIT (normalized to the lazy scheme).}
	\label{cachehitratio}
	\vspace{-0.3cm}
\end{figure}

As shown in Fig.~\ref{cachehitratio}, the lazy and eager schemes in SIT exhibit an approximate number of metadata cache misses (the experimental configurations are shown in Section~\ref{evaluation}). However, since the lazy scheme doesn't update the root~\cite{ZubairA19}, we focus on the eager scheme to reduce the root update latency.

In this paper, in order to improve the system performance, we propose SCUE, a low-overhead shortcut update scheme, to immediately update the root, ensure the root crash consistency and provide the ability to reconstruct the SIT from leaf nodes for verification after system reboots.

\section{System Design and Implementations}
\label{design}

\begin{figure}[t]
	\vspace{-0.3cm}
	\centering
	\includegraphics[width=0.45\textwidth]{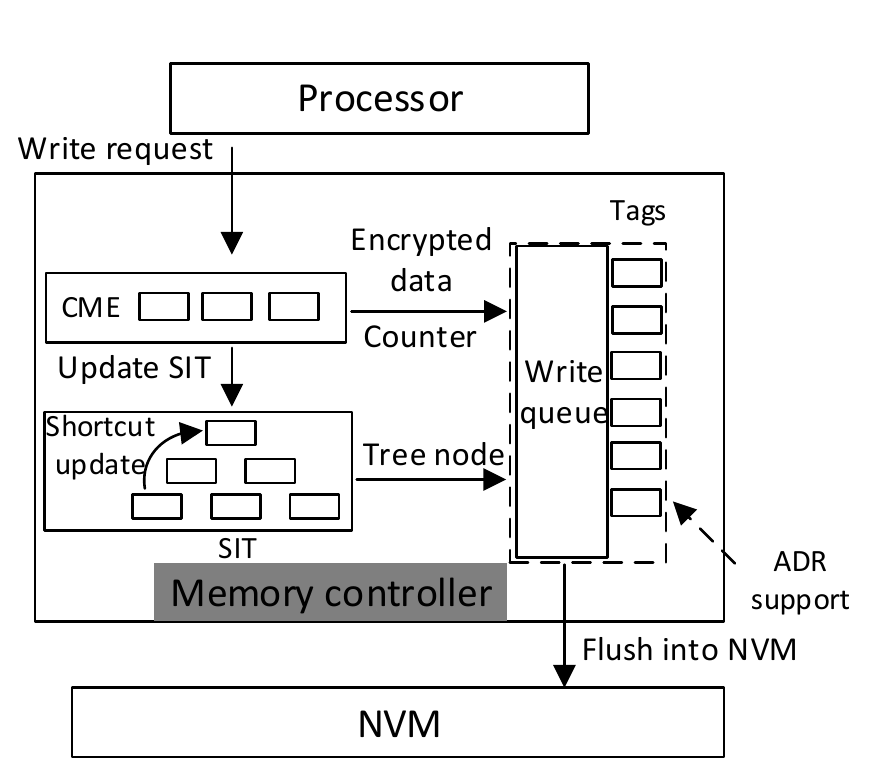}
	\vspace{-0.1cm}
	\caption{The hardware architecture of SCUE.}
	\label{overview}
	\vspace{-0.3cm}
\end{figure}

\subsection{Overview}
In the secure NVM systems, the encryption and integrity verification are performed in the memory controller and transparent to applications. As shown in Fig.~\ref{overview}, a processor sends a write request to the memory controller. The written data is encrypted by the CME and updates the SIT. The data then enters the write queue supported by ADR. To ensure the consistency of data, counter and tree node, existing schemes offer different approaches to flushing the counters and tree nodes into the write queue~\cite{YangLCMS19,ZubairA19,LiuKRK18}, which unfortunately fail to offer efficient update upon the root in SIT. To support the verification after system reboots, the systems also need to ensure the consistency between the root and the leaf nodes.

We propose the SCUE to significantly reduce the latency of updating tree root and ensure the root crash consistency. As shown in Fig.~\ref{overview}, modifying the counter blocks in CME causes the changes in the entire SIT, and we overlook the intermediate nodes to directly update the root of SIT via SCUE since the updates in the intermediate tree nodes are unnecessary. The tags are appended on the write queue and used to store the updated roots. The counter crash consistency~\cite{zuo2019supermem,LiuKRK18,YeHA18} is beyond the scope of this paper, and we use Osiris~\cite{YeHA18} to ensure the counter blocks are up-to-date and consistent. The encrypted user data contain the updates of the counter blocks as the leaf nodes in the SIT. Due to the use of ADR, the encrypted data are flushed into the NVM while the corresponding roots are flushed into the non-volatile register on chip to ensure the root crash consistency.

\begin{figure}[t]
	\vspace{-0.3cm}
	\centering
	\includegraphics[width=0.45\textwidth]{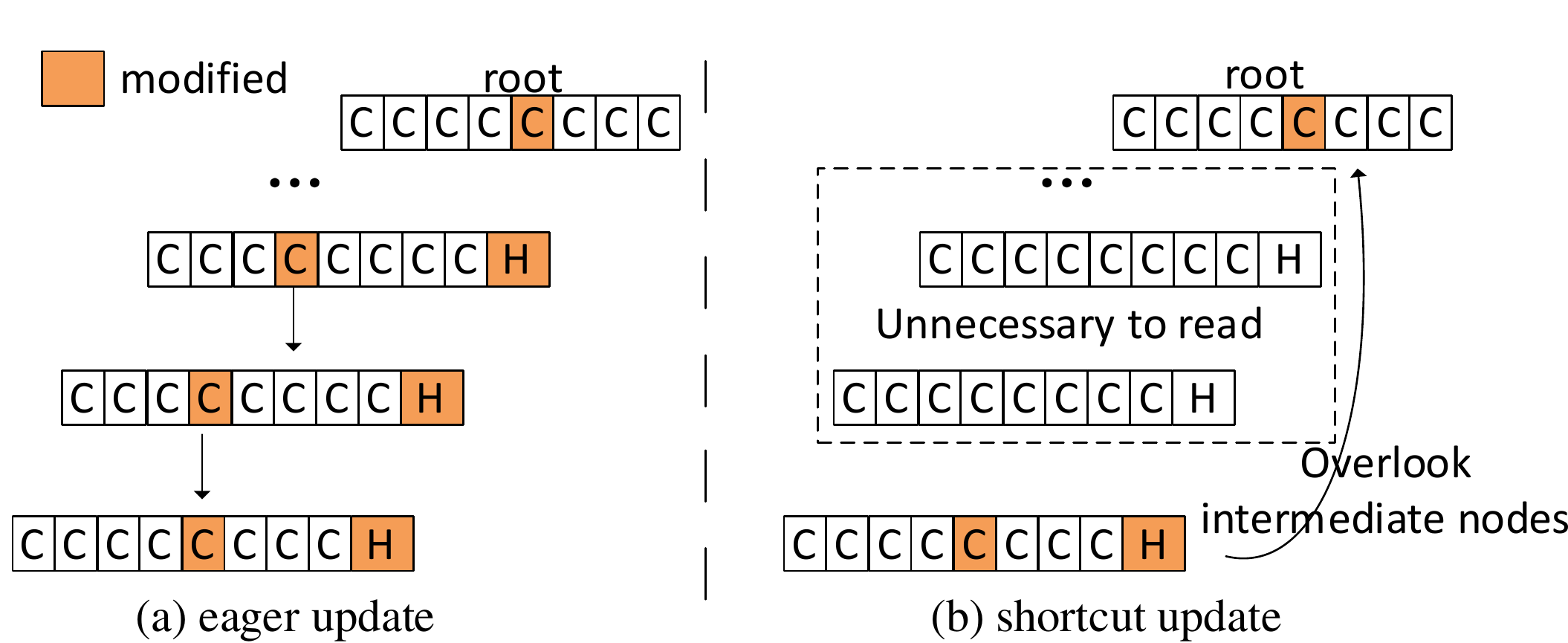}
	\vspace{-0.1cm}
	\caption{The eager and SCUE schemes in SIT. (a) Eager update scheme modifies the intermediate nodes. (b) SCUE overlooks the intermediate nodes.}
	\label{eands}
	\vspace{-0.5cm}
\end{figure}

\subsection{Update the Root in a Shortcut Manner}
\label{shortcut}
To reduce the latency of updating the root of SIT eager scheme and improve system performance, we propose \textit{a low-overhead shortcut update (SCUE) scheme}. The basic idea of SCUE is to remove the computation of HMACs and reads of intermediate tree nodes from the write critical path.

\subsubsection{Lazy computing in SCUE}

We observed that in SIT nodes, the HMACs are used to verify the node itself, and the counters are used to verify the child nodes. When reading data from NVM into cache, the counters in the cached tree nodes and HMACs in the read data are used to verify the integrity of the read data. Since the on-chip data are invulnerable to attackers in our threat model, the integrity of the cached tree nodes needn't be verified, and the HMACs are not used. To verify the integrity of the data next time they are read, the counters in cached tree nodes need to increase when the data are persisted, but updating HMACs is unnecessary. Only the HMACs in the persisted data need to be updated to defend against attacks that occur in NVM.

We propose lazy computing in SCUE to remove the long hash\_latency in the propagation path from leaf nodes to root. When one user data is persisted, the HMAC in the user data is computed to facilitate the verification of the following read. The counters in the ancestor tree nodes increase, but their HMACs are only computed when the nodes are flushed into NVM via the cache replacement policy.

\subsubsection{Remove the reads of tree nodes}
Propagating the modifications from leaf nodes to root needs to read the ancestor nodes with long read\_latency, if the ancestor nodes are not in cache. In SIT, the root consists of eight counters, and the counter values are related to the leaf nodes. As shown in Fig.~\ref{eands}(a), in the eager update scheme, when one modified leaf node is persisted, its ancestor nodes are updated. Specifically, all corresponding counters increase by one, including the counter in root. In summary, if one modified leaf node is persisted into NVM, the corresponding counter in root increases by one. To improve the system performance, as shown in Fig.~\ref{eands}(b), we propose the SCUE to directly increase the counter in root without reading the intermediate tree nodes.

\begin{figure}[t]
	\vspace{-0.3cm}
	\centering
	\includegraphics[width=0.25\textwidth]{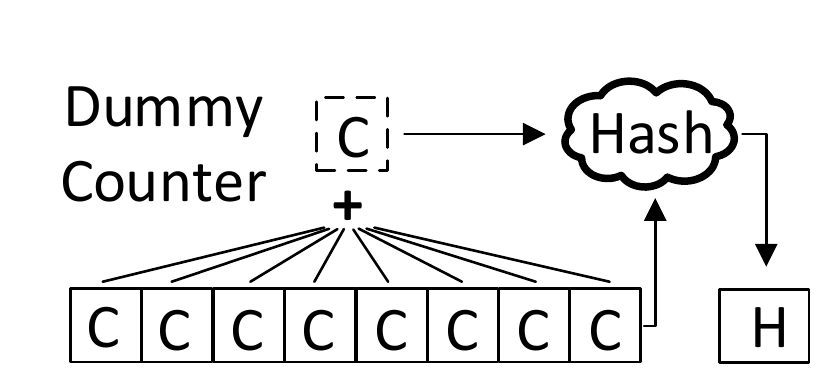}
	\vspace{-0.1cm}
	\caption{Dummy counter is generated by summing all counters in the tree node.}
	\label{dummycounter}
	\vspace{-0.5cm}
\end{figure}

However, the root update by overlooking intermediate nodes needs to carefully handle the system execution. When the intermediate nodes are read, they should be iteratively verified by the cached tree nodes and root. However, the updated root can't match the un-updated intermediate nodes since we overlook the intermediate nodes to update the root. Therefore, the attacks are misreported even if the intermediate nodes are un-attacked. To verify the data integrity on running time, we read and update the intermediate tree nodes via lazy computing in the backend threads without computing HMACs after the root is updated. In SCUE, after updating the root counters, the write operations are completed. The latencies of reading the intermediate tree nodes are removed from the write critical path.

Moreover, when the cached tree nodes are flushed into NVM due to cache replacement policy, their parent nodes need to be read to compute the HMACs of the persisted nodes, which possibly incurs iterative reads of ancestor nodes to execute the integrity verification. To reduce these reads, SCUE proposes the \textit{dummy counter} as shown in Fig.~\ref{dummycounter}. A dummy counter is generated by summing all counters in the tree node being flushed. In the eager scheme, when any counter in one node increases by one, the corresponding counter in the parent node (called\textit{ parent counter}) also increases by one. Therefore, the dummy counter becomes the parent counter of the persisted node to compute the HAMC. Since the cached node is safe in our threat model, the generated dummy counter is the same as the parent counter. The iterative reads of the ancestor nodes are hence removed.

\subsubsection{Benefits of the proposed optimizations}

Persisting user data causes the modifications of the counter blocks, which are the leaf nodes of the SIT. SCUE propagates the modifications by directly updating the root without reading and updating intermediate nodes. The reads and updates of the intermediate nodes are executed via lazy computing by the backend threads. When the cached tree nodes are flushed into NVM, the dummy counters are generated to compute the HMACs if the parent nodes of the flushed nodes are not in cache. Even compared with lazy scheme which doesn't update the root, SCUE also delivers better performance, since lazy scheme needs to compute the HMACs in the written data and parent node and iteratively read the ancestor nodes when one data is persisted.

In summary, for writing data, SCUE removes all reads of integrity tree nodes from the write critical path and only needs to generate one HMAC in the persisted data.

\subsection{Reconstruct the SIT}
SIT fails to be reconstructed from the leaf nodes ~\cite{ZubairA19,alwadi2019phoenix,freij2020streamlining} since reconstructing one SIT node requires its parent node as input. We leverage a counter-summing approach to reconstructing the SIT from leaf nodes up and analyze the security of the proposed scheme.  

\subsubsection{Counter-summing approach to reconstructing tree}
 
In SIT, even if the root is timely updated via SCUE, due to failing to reconstruct the integrity tree from the leaf nodes, the up-to-date root in SIT fails to verify the leaf nodes after system reboots. In existing integrity tree schemes, only the eager updated BMT and MT, which can be reconstructed from leaves and contain up-to-date root, offer integrity verification after system failures with a long root update latency during running time. To provide efficient integrity verification for SCUE, we use the counter-summing recovery approach to reconstructing the SIT.

In SIT, each node contains eight counters and has eight child nodes. We observed that when one child node is modified, i.e., any counter increases by one, the corresponding counters in the parent and ancestor nodes also increases by one. As shown in Fig.~\ref{dummycounter}, the dummy counter is generated according to the observation. After the system reboots, we reconstruct the tree from the bottom up via the dummy counter.

We first reconstruct the Level-1 nodes from the consistent leaf nodes. Reconstructing one SIT node requires correct counters and HMAC in the node. For reconstructing counters, we generate the dummy counters from the leaf nodes, and each $8K$th to $8K+7$th ($K$ is a natural number) dummy counters are combined to form one Level-1 node. The dummy counters become counters in the reconstructed nodes. We then use the HMACs in the persistent leaf nodes to verify the correctness of the reconstructed counters in Level-1 nodes. The HMAC is recomputed by hashing the reconstructed counter in the Level-1 node and all counters in the corresponding leaf node. The mismatches of recomputed HMACs and stored HMACs in the leaf nodes indicate attacks occurring during system recovery. All counters in the Level-1 nodes can be recovered via summing the corresponding counters in the leaf nodes.

For recomputing HMACs, the counters in Level-2 nodes need to be reconstructed by generating the dummy counters from Level-1 nodes. The HMACs in the Level-1 nodes are recomputed by hashing the corresponding counters in the Level-2 nodes and all counters in the Level-1 nodes. Therefore, the counters and HMACs in the Level-1 nodes are reconstructed. We use the same way to reconstruct the whole tree from the bottom up. Finally, the dummy root counters are generated. If the dummy root counters are different from the stored root counters, the SIT reconstruction fails, and attacks occur during recovery. Otherwise, the SIT is successfully reconstructed.

\subsubsection{Security analysis}
We analyze the security of the proposed counter-summing recovery approach of SIT. We focus on the root and leaf nodes since the intermediate tree nodes are lost when failures occur, which need to be reconstructed from the persistent leaf nodes. Moreover, since the root is on chip, attackers can't tamper with the root. When recovering the SIT, only the leaf nodes in NVM are attacked. 

\begin{table}[!tbp]
	\vspace{-0.2cm}
	\footnotesize
	\caption{\label{table:security}The attacks can be detected by HMACs and root.}
	\vspace{-0.5cm}
	\begin{center}
		\begin{tabular}{|c|c|c|c|}
			\hline
			&\tabincell{c}{Roll-forward\\ attacks}& \tabincell{c}{Roll-back\\ attacks}  & \tabincell{c}{Roll-forward\\ + roll-back attacks}     \\
			\hline
			HMACs in leaf nodes & detected &  & detected              \\
			\hline
			Root &  & detected &              \\
			\hline
		\end{tabular}
	\end{center}
	\vspace{-0.5cm}
	
\end{table}

We use the HMACs in the leaf nodes and the on-chip root to detect the attacks that occur during system recovery. All attacks in the leaf nodes can be divided into two types: roll-forward and roll-back attacks. Specifically, the roll-forward attacks tamper with the counter value in the leaf nodes to a larger value. On the contrary, the roll-back attacks tamper with the counter value in the leaf nodes to a smaller value. The replay attack is a typical roll-back attack that replaces the counters and HMACs using the old tuple.

As shown in Table~\ref{table:security}, the roll-forward attacks are detected by the HMACs in the leaf nodes. The parent counters are reconstructed from the leaf nodes, and the HMACs of the leaf nodes are recomputed. Without the secret key, attackers can't calculate the correct HMACs using new counters. The mismatches between the recomputed HMACs and stored HMACs in leaf nodes indicate the attacks. 

The normal roll-back attacks are also detected by HMAC since the HMAC doesn't match the tampered counters. We discuss how to detect the replay attacks, which pass the verification of HMAC since the old HMAC matches the old counters. Since increasing counters in leaf nodes causes the increment of counters in root, one root counter is equal to the sum of all counters in the corresponding leaf nodes. For example, the value of first counter in the root is the sum of counters' values in the leading one eighth of the nodes in the leaf level since the root contains eight counters. Rolling back counters in leaf nodes causes the reconstructed root to mismatch the stored root. Therefore, the attacks are detected. Moreover, attackers may roll back and roll forward some counters at the same time to deceive the root, but rolling forward counters is detected by the HMACs as shown in Table~\ref{table:security}. The detection process is similar to that of BMT, i.e., the attacks are detected after reconstructing the whole integrity tree and comparing the reconstructed root with the stored one.

The counter-summing recovery approach allows to reconstruct the SIT from consistent leaf nodes. Therefore, the SIT root can be used to verify data after system failures. The related tree works~\cite{freij2020streamlining,AwadYSNZ19} for better system performance can be used by SIT.

\begin{figure}[t]
	\vspace{-0.3cm}
	\centering
	\includegraphics[width=0.45\textwidth]{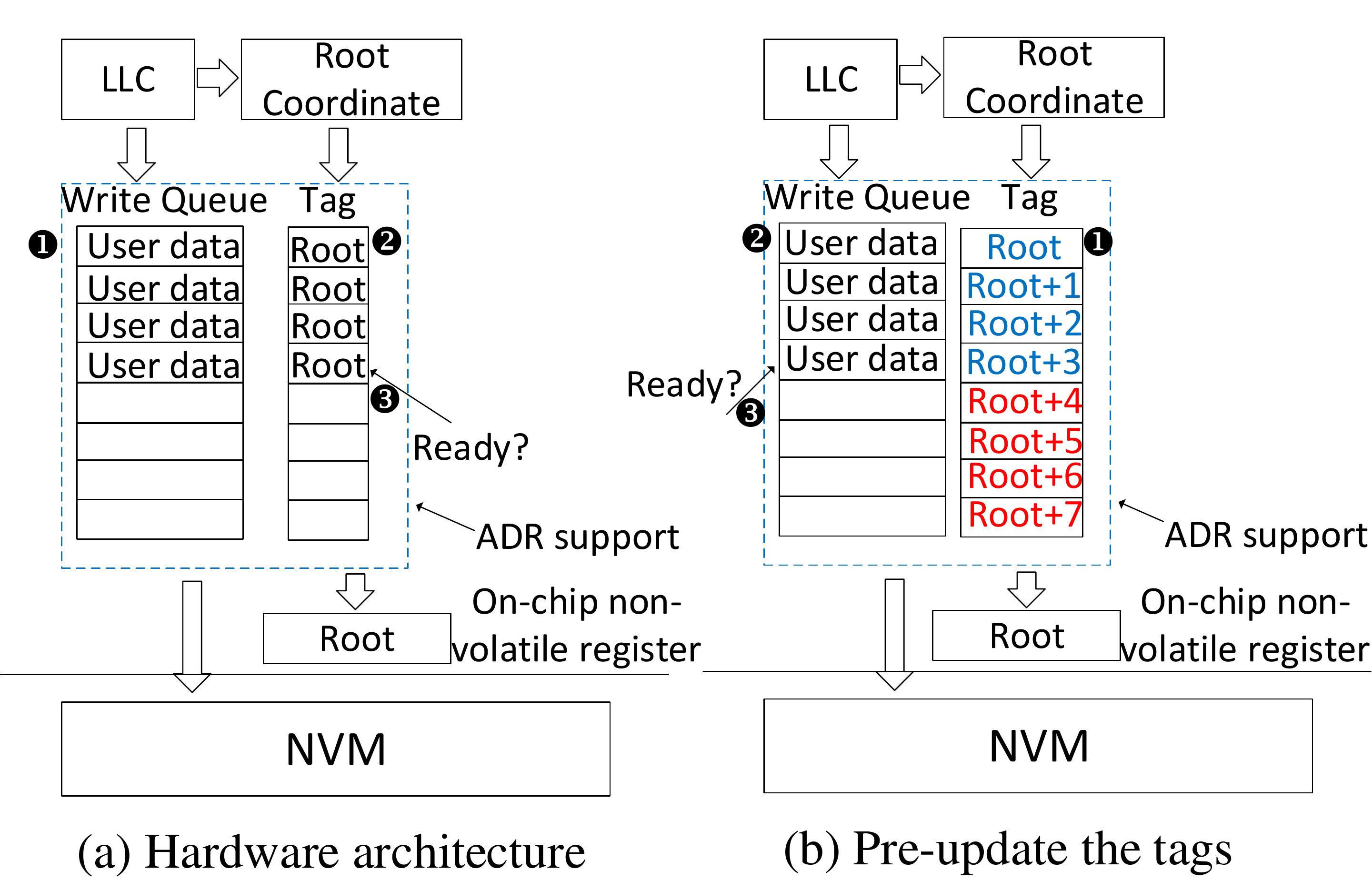}
	\vspace{-0.1cm}
	\caption{Hardware implementations. (a) Hardware architecture for root crash consistency. (b) Pre-update the tags for concurrent threads. }
	\label{EarlyUpdate}
	\vspace{-0.3cm}
\end{figure}

\subsection{Ensure Root Crash Consistency}
Like SCA~\cite{LiuKRK18}, we slightly modify the write queue to ensure the crash consistency between root and leaf nodes. However, the intuitive design fails to be used in a modern multi-thread platform. We pre-update the root before leaf nodes are modified to address the problem of root crash inconsistency.

\subsubsection{Intuitive hardware design}

SCUE updates the root of SIT with negligible latency. However, persisting the leaf nodes and updating the root are still two atomic operations. When system failures occur between these two operations, the root is inconsistent with the leaf nodes and can't be used to verify the leaf nodes after system reboots. To ensure the root crash consistency, we slightly modify the hardware architecture like the SCA~\cite{LiuKRK18}. As shown in Fig.~\ref{EarlyUpdate}(a), the write queue is supported by ADR. Once power off, the data in the write queue can be flushed into NVM with the support of the backup battery. To persist the root, we extend the write queue in the memory controller, i.e., each entry in the write queue has a tag, in which we store the corresponding root. Since we use Osiris~\cite{YeHA18} to ensure the counter crash consistency, the user data contain the updates of the counter blocks. Persisting user data means that the counter blocks are persisted, which are the leaf nodes in the SIT. When the user data enter the write queue (\ding{182}), we fill the tags using the roots (\ding{183}), which are updated by increasing the corresponding counter via SCUE. When the tags become ready (\ding{184}), the user data in the write queue are allowed to flush into NVM, and the roots in tags are also flushed into the on-chip root register.

We analyze the root crash consistency in the intuitive design. If the system failures occur before the tags are prepared, the new user data are not allowed to be flushed into NVM. The old root can match the old and consistent leaf nodes in NVM when the leaf nodes are recovered from user data via Osiris. If the system failures occur after the tags are prepared, the user data in the write queue are flushed into NVM due to the ADR support, and the updated root in the tag is also flushed into the on-chip root register. Thus both the root and leaf nodes become new states. 
\subsubsection{Pre-update approach}

The intuitive hardware design is inspired by SCA~\cite{LiuKRK18} and can work well on the single-thread architectures. However, in the modern multi-thread architectures, the root fails to be correctly updated. When user data concurrently enter the write queue, multiple threads read the root and increase the corresponding counter value by one to fill the tags. The multiple tags are incorrectly updated to the same values since the same root is read and root counters increase by one by the threads.

To correctly update the roots in tags, we observe that the updates of root counters are predictable, i.e., the counters of roots in consecutive tags are incremented in most cases. Since one counter in root covers 1/8 memory space, persisting two data may cause the same counter in root modified, i.e., the two data share the same root counter. Assume one counter in root is $n$, persisting one user data changes the counter to $n+1$ in the root. Following user data persistence also increases this counter to $n+2$ if the user data shares the same root counter with the previous data. Based on this observation, we propose a \textit{Pre-update} approach to filling the tags in advance. As shown in Fig.~\ref{EarlyUpdate}(b), the background thread writes $Root$ to the 1st tag and $Root+1$ to the 2nd tag. The remaining tags are also prepared by increasing the value in the previous tag(\ding{182}). The write queue concurrently receives user data(\ding{183}) as filling the tags.

When pre-updating the tags in the write queue, we need to determine which counter in root needs to increase by one. The root contains eight counters, and the memory is equally divided into eight parts. According to the location of the user data, persisting the data increases one corresponding counter in the root. If the user data entering the write queue makes another counter, which is not the counter prepared in advance, in the root modified, the prepared tags need to be cleared and a new root fills the tag. The write queue is blocked until the new tags are prepared for the root crash consistency.

In general, applications have high spatial locality and one counter in root covers 1/8 memory space. It is a high probability that the following persisted user data share the same root counter as the previous persisted user data. Thus we wait for the root counter to be modified and increase the same root counter in the consecutive tags in advance. If persisting one data needs to increase another root counter in the tag, we re-update this tag and remaining tags.

\begin{figure}[t]
	\vspace{-0.3cm}
	\centering
	\includegraphics[width=0.45\textwidth]{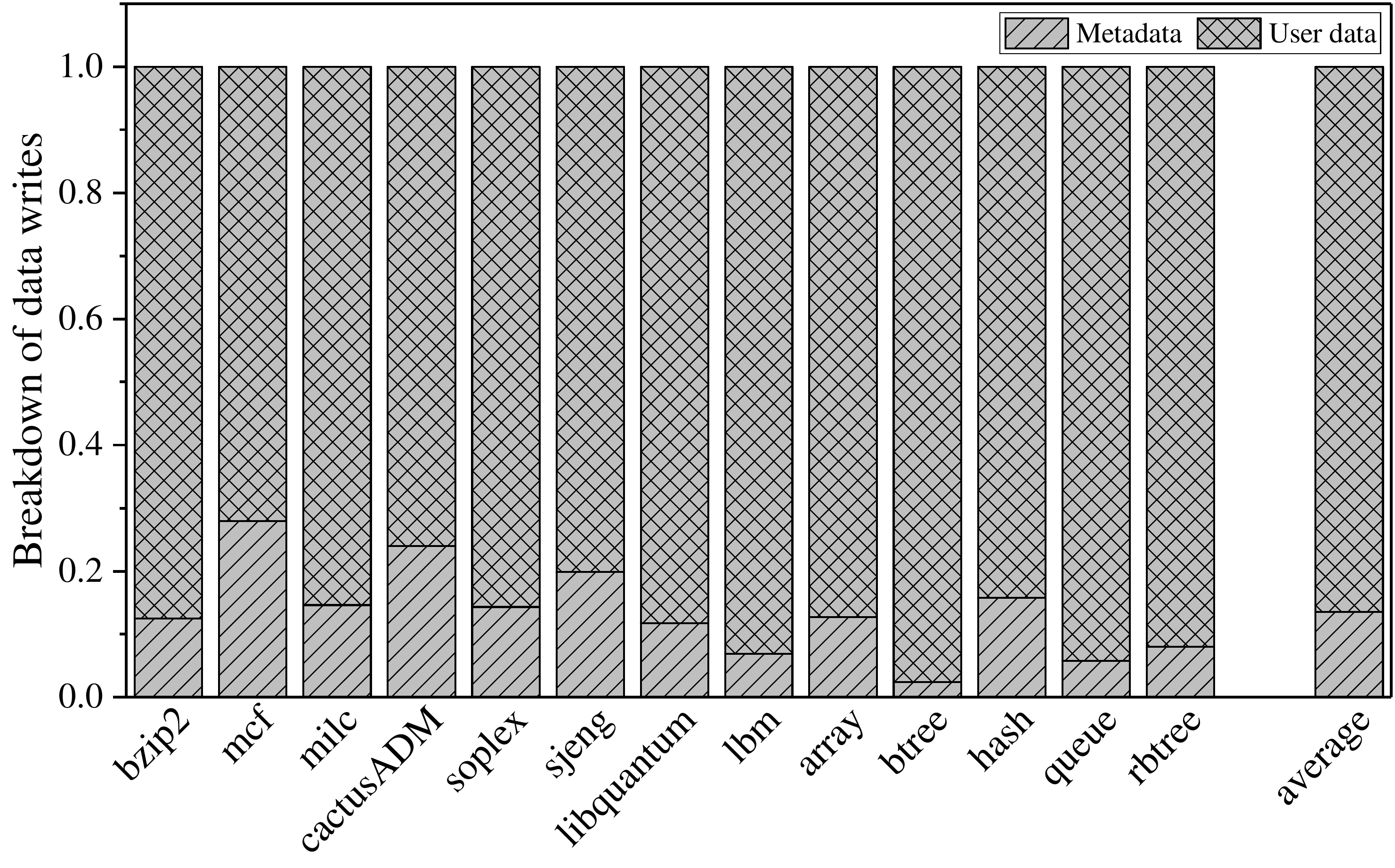}
	\vspace{-0.3cm}
	\caption{The breakdown of data writes in SCUE.}
	\label{metadata_ratio}
	\vspace{-0.3cm}
\end{figure}

The tags in the write queue are used to store the counterparts of the 64-byte root. An entry in the write queue is also 64 bytes. Thus the write queue needs to extend 1x space to store the tags, which consumes the expensive ADR space. However, only one counter in root increases when one user data enters the write queue. To reduce the space overhead, we store the modified 64-bit counter in the tag, instead of the full root.

When system failures occur, due to the support of ADR, the data in the write queue are flushed into NVM. If the root counters in tags have corresponding data in the queue, the root counters are flushed into the on-chip non-volatile register to update the root. As shown in Fig.~\ref{EarlyUpdate}(b), the blue root counters are flushed into the register since their corresponding user data are flushed into NVM, while the red root counters can't be flushed. Flushing the root counters is simple. Due to the first-in-first-out feature of the write queue, the old root counter is first flushed. The latest root counter finally covers the old root counter in the register, so that the root becomes the latest to match the leaf nodes that are recovered from the persistent user data.

Persisting tree nodes doesn't impact the root in SCUE. However, when entering the write queue, the tree nodes have the corresponding tags. After system failures, the root is incorrectly updated via ADR. Data writes contain security metadata and user data writes. Fig.~\ref{metadata_ratio} shows the breakdown of data writes. On average, the amount of metadata writes is 13.6\% of all data writes. For the 64-entry write queue~\cite{wei2020morlog}, we add another 10 entries without tags to handle metadata writes like SCA~\cite{LiuKRK18}. Persisting tree nodes thus doesn't incorrectly increase counters in the root.

\vspace{-0.2cm}
\subsection{Comparisons of Integrity Tree Update Schemes}

\begin{table}[!tbp]
	\footnotesize
	\caption{\label{table:comparisons}The comparisons of different integrity tree update schemes.}
	\vspace{-0.3cm}
	\begin{center}
		\begin{tabular}{|c|c|c|c|c|c|}
			\hline
			 &\tabincell{c}{Lazy \\ BMT} &\tabincell{c}{Eager \\ BMT} &\tabincell{c}{Lazy \\ SIT} &\tabincell{c}{Eager \\ SIT} & \tabincell{c}{SCUE \\ for SIT}    \\
			\hline
			\tabincell{c}{Latency of\\ updating root} & / & high & / & normal & negligible             \\
			\hline
			\tabincell{c}{Recoverability} & N & Y & N & N & Y            \\
			\hline
			\tabincell{c}{Root crash\\ consistency} & N & N & N & N & Y        \\
			\hline
		\end{tabular}
	\end{center}
	\vspace{-0.3cm}
	
\end{table}

We compare existing schemes of updating integrity trees with our proposed SCUE. As shown in Table~\ref{table:comparisons}, the eager BMT requires long latency to update the root. The eager SIT also needs to propagate the modifications, and however, the eager SIT uses multiple hash circuits to update the intermediate nodes in parallel with high computation overhead in the memory controller. Thus the latency of updating eager SIT root is shorter than that of eager BMT root. Our SCUE needs a negligible latency to update root, since the scheme overlooks the intermediate nodes and directly updates the root. The lazy BMT/SIT doesn't update the tree roots. They can't recover after failure due to the loss of the latest root. Moreover, the eager SIT can't be reconstructed from leaf nodes due to the unique dependency between the counters in parent nodes and HMACs in the child nodes.

In existing works, we only reconstruct the eager BMT from the leaf nodes during recovery. However, the long root update latency and root crash inconsistency problems are not considered by existing integrity tree update schemes. Our proposed SCUE is able to immediately update the root of SIT, recover the SIT from the leaf nodes and ensure the root crash consistency.

\section{Performance Evaluation}
\label{evaluation}


\subsection{Evaluation Methodology}
\label{section 5}
To evaluate the performance of SCUE, we model the system in the Gem5~\cite{BinkertBBRSBHHKSSSSVHW11} and NVMain~\cite{PorembaZ015}. NVMain is a cycle-accurate main memory simulator for emerging NVM technologies. Main parameters are shown in Table~\ref{table:configure}. The metadata cache in the memory controller is 256KB, storing counter blocks and integrity tree nodes. We model the PCM technologies with 16GB capacity. The PCM latency is modeled like existing works~\cite{zuo2019supermem,xu2015overcoming}. In general, the hash latency to generate the HMAC is 80 cycles~\cite{gassend2003caches}. In our sensitive study, we vary the hash latencies (from 80 to 160 cycles~\cite{liu2019janus,suh2003efficient}) to analyze the impacts on performance.
We use 8 representative applications from CPU SPEC2006 benchmarks~\cite{Henning06}, simulating 5 billion instructions for each application, and 5 persistent workloads to evaluate the systems. The persistent workloads, i.e., array, btree, hash, queue and rbtree, are widely used in existing works of persistent memory~\cite{CoburnCAGGJS11,RenZKCWM15,KolliRDSPLCW16,KolliGSDCNW17,LiuKRK18,zuo2019supermem}.

\begin{table}[!tbp]
	\footnotesize
	\caption{\label{table:configure}The configurations of the evaluated NVM system.}
	\vspace{-0.3cm}
	\begin{center}
		\begin{tabular}{|c|l|}
			\hline
			\multicolumn{2}{|c|}{\textbf{Processor}} \\
			\hline
			CPU & 8 cores, X86-64 processor, 2 GHz        \\
			\hline
			Private L1 cache & 64KB, 2-way, LRU, 64B Block              \\
			\hline
			Private L2 cache & 512KB, 8-way, LRU, 64B Block        \\
			\hline
			Shared L3 cache & 4MB, 8-way, LRU, 64B Block            \\
			\hline
			
			\multicolumn{2}{|c|}{\textbf{DDR-based PCM Main Memory}} \\
			\hline
			Capacity &  16GB     \\
			\hline
			PCM latency model &   \tabincell{c}{tRCD/tCL/tCWD/tFAW/tWTR/tWR \\=48/15/13/50/7.5/300 ns}     \\
			\hline
			Write queue & \tabincell{c}{64 entries with 64-bit tags for user data, \\ 10 entries without tags for security metadata}   \\
			\hline
			\multicolumn{2}{|c|}{\textbf{Secure Parameters}} \\
			\hline
			Security metadata cache & \tabincell{c}{256KB, 8-way, 64B Block, in MC}  \\
			\hline
			SIT & \tabincell{c}{9 levels, 8-ary, 64B Block}             \\
			\hline
			Hash latency & \tabincell{c}{80 cycles~\cite{gassend2003caches}}             \\
			\hline
		\end{tabular}
	\end{center}
	\vspace{-0.2cm}
	
\end{table}

To comprehensively examine the performance of our proposed SCUE, we evaluate and compare the following schemes.
\vspace{-0.1cm}
\begin{itemize}
	\setlength{\itemsep}{0pt}
	\setlength{\parsep}{0pt}
	\setlength{\parskip}{0pt}
	\item Eager scheme (Eager). An eager update scheme propagates the modifications from the modified leaf nodes to the root with a large overhead. The BMT-eager scheme verifies the leaf nodes after system failures since the integrity tree can be reconstructed from the persistent leaf nodes.
	\item Lazy scheme (Lazy). The lazy scheme updates the parent nodes of the written data in the integrity tree but doesn't propagate the modifications to the root. It also needs to read the ancestor nodes to verify the parent node when writing data. The lazy scheme doesn't offer integrity verification after system failures and reboots. 
	\item Our proposed lazy computing (LC). LC increases the counters of the tree nodes in the propagation path. Only the HMACs in the persisted data are computed. 
	\item Shortcut update scheme (SCUE). SCUE directly updates the root without propagating the modifications from leaf nodes to root. In SCUE, the HMACs are calculated via LC, and the intermediate tree nodes are not read on the write critical path.

\end{itemize}

\vspace{-0.3cm}
\subsection {Results and Analysis}
Propagating the modifications from leaf nodes to the root incurs a long latency on the write critical path. We evaluate the write latency and execution time on different schemes.

\begin{figure}[t]
	\centering
	\includegraphics[width=0.45\textwidth]{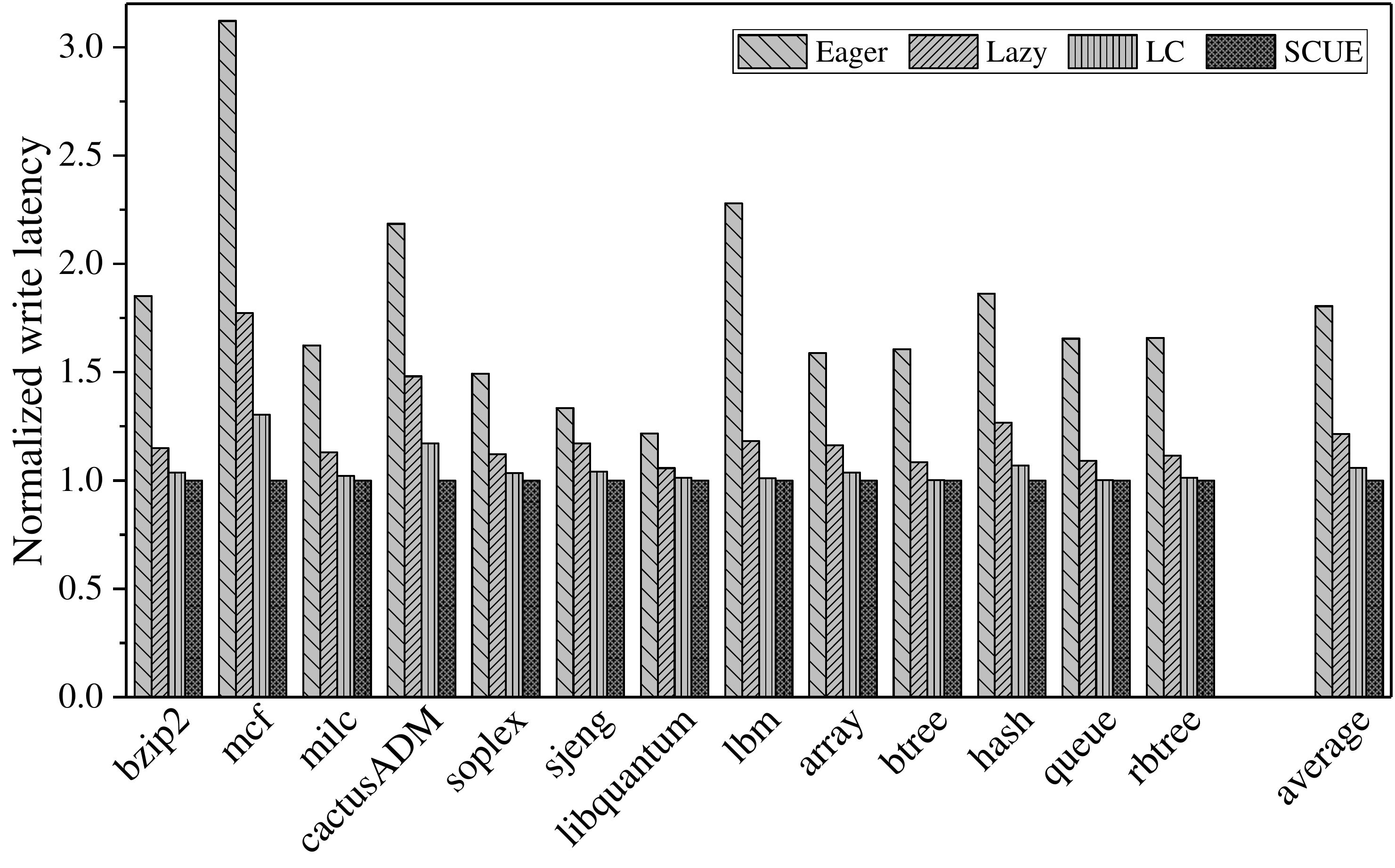}
	\vspace{-0.3cm}
	\caption{The write latency on different workloads (normalized to SCUE).}
	\label{write_latency}
	\vspace{-0.3cm}
\end{figure}

\begin{figure}[t]
	\centering
	\includegraphics[width=0.45\textwidth]{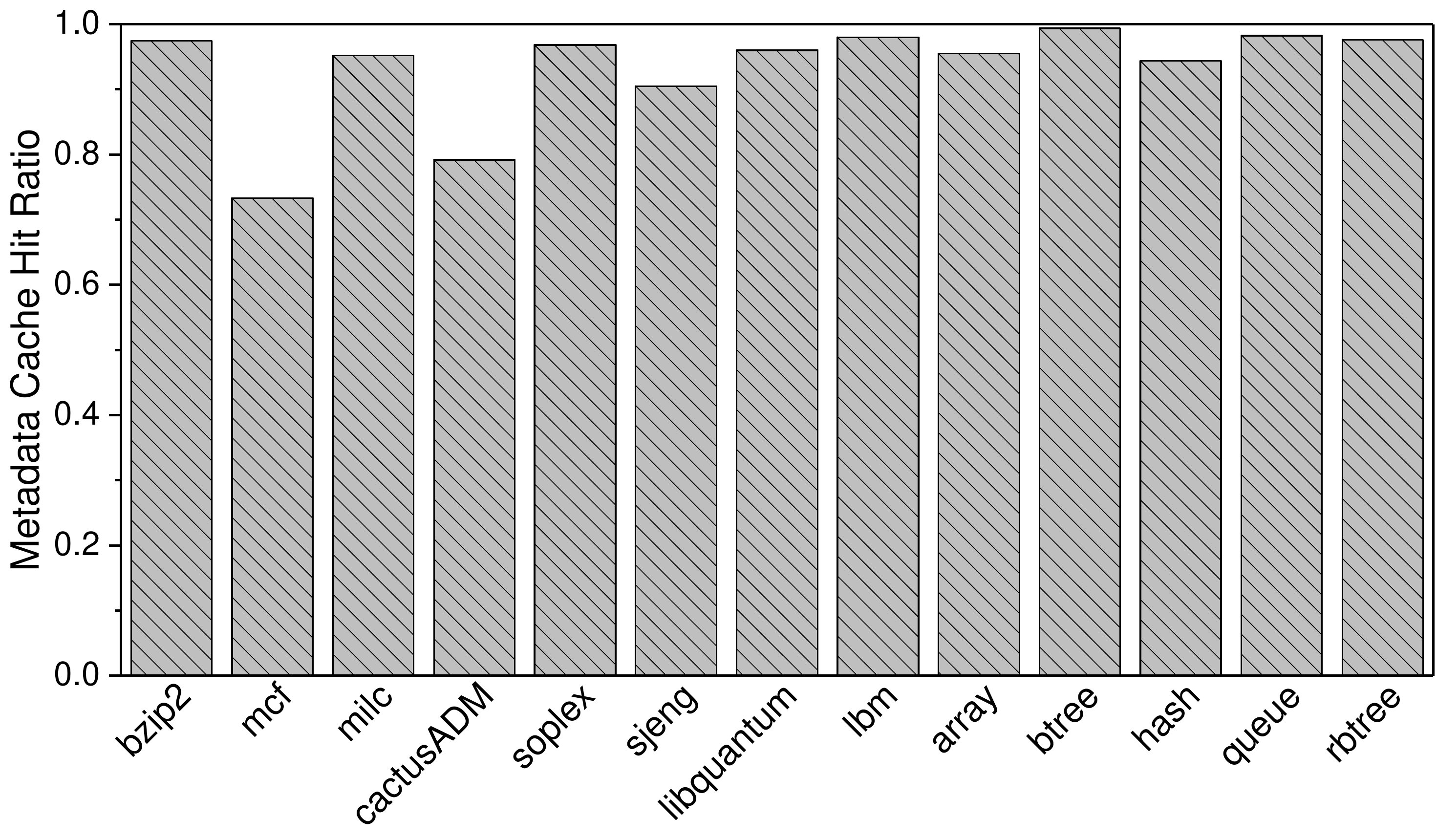}
	\vspace{-0.3cm}
	\caption{The metadata cache hit ratio on different workloads during the write process.}
	\label{write_hitratio}
	\vspace{-0.5cm}
\end{figure}

\begin{figure}[t]
	\centering
	\includegraphics[width=0.45\textwidth]{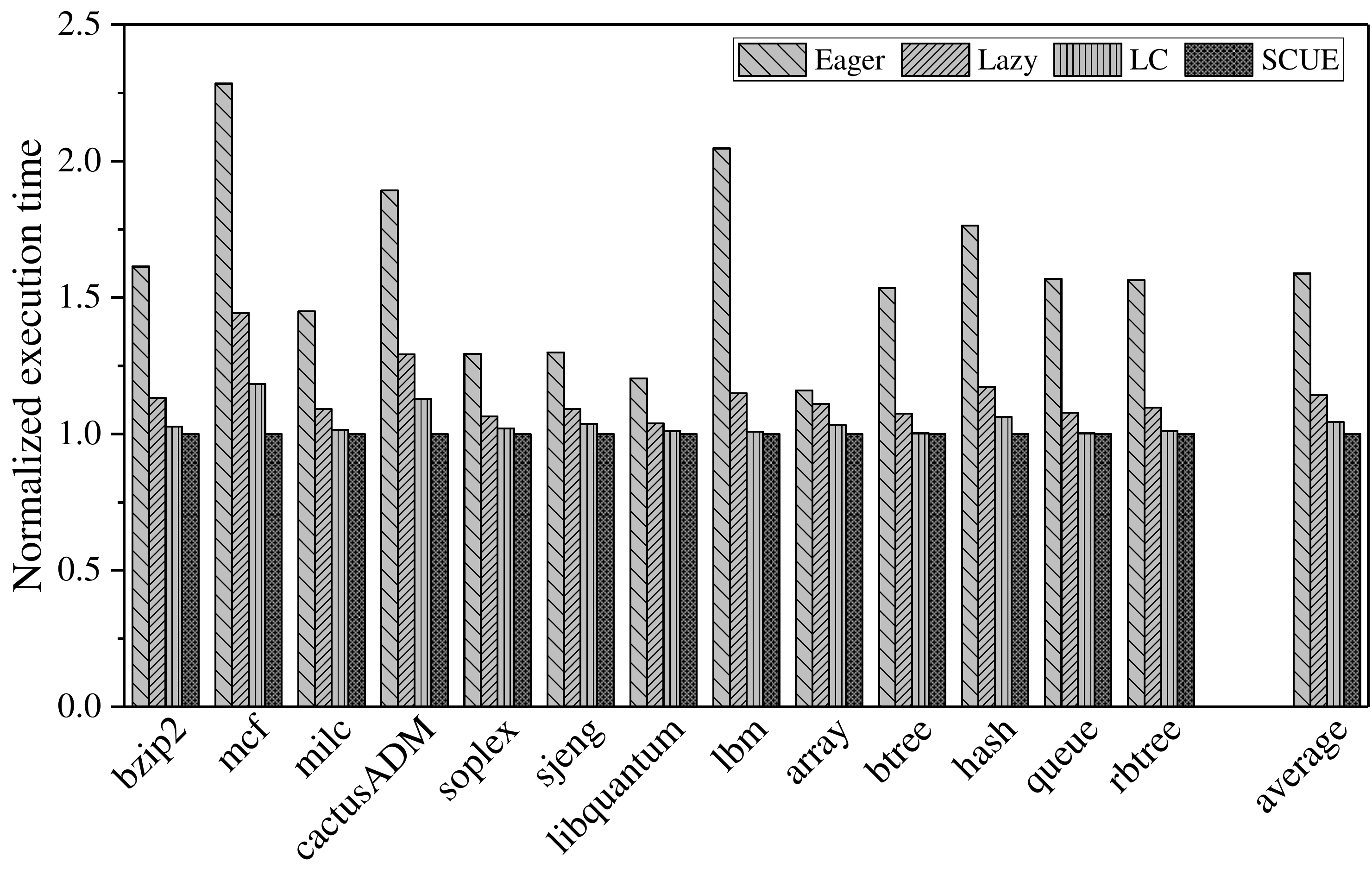}
	\vspace{-0.3cm}
	\caption{The execution time on different workloads (normalized to SCUE).}
	\label{execution_time}
	\vspace{-0.3cm}
\end{figure}

Fig.~\ref{write_latency} shows the write latencies on different schemes. Due to performing multiple hash calculations before persisting data, on average, the write latency in the Eager scheme is 1.81x than that in SCUE. Moreover, the Lazy scheme needs to read the parent and ancestor nodes of the evicted user data and metadata, and calculate the HMACs in the evicted data and parent nodes on the write critical path. Our SCUE needn't read any nodes and only compute the HMAC once when writing data. Therefore, SCUE has a lower write latency than the Lazy scheme. On average, the write latency in Lazy scheme is 1.21x than that in SCUE.

We also compare LC with the SCUE. SCUE contains the LC approach but removes the data reads of the ancestor nodes. Compared with the LC approach, the performance improvement of SCUE is related with the number of metadata reads from NVM when writing data. Since many metadata are cached during the data read process, Fig.~\ref{write_hitratio} shows that the metadata cache hit ratio is high during the write process. Many ancestor nodes already exist in cache and needn't be read from NVM. Therefore, SCUE shows a light improvement (5\% on average) on write latency than the LC approach. However, for the workloads with low metadata cache hit ratios, such as mcf and cactusADM, SCUE shows much lower write latency than the LC approach when using the high hit ratio workloads, such as lbm, btree and queue.

As shown in Fig.~\ref{execution_time}, SCUE shows a lower execution time than the Eager and Lazy schemes. Specifically, the Eager scheme leads to a 1.59x slowdown than SCUE. On the write-intensive workloads, such as mcf, the slowdown of the Eager scheme is up to 2.28x. SCUE delivers a higher performance than the Lazy scheme in terms of execution time. The average execution time in the Lazy scheme is 1.14x than that in SCUE, while SCUE is able to verify the integrity after system reboots and the Lazy scheme fails. Moreover, SCUE achieves less execution time than LC as shown in Fig.~\ref{execution_time}. Like the write latency in Fig.~\ref{write_latency}, compared with LC, SCUE shows higher performance improvement when using lower hit ratio workloads.

\begin{figure}[t]
	\centering
	\includegraphics[width=0.45\textwidth]{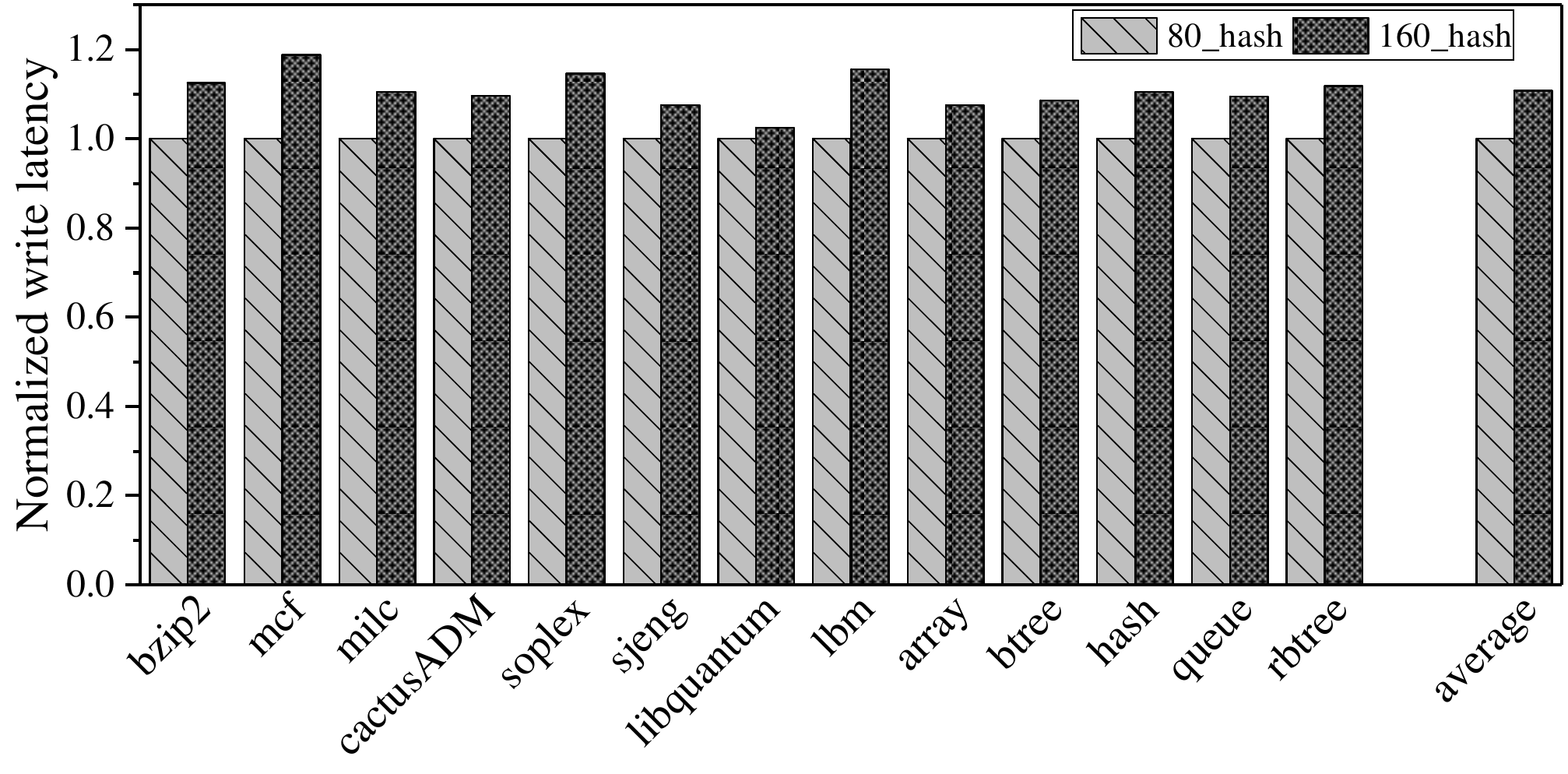}
	\vspace{-0.1cm}
	\caption{The write latency of using different hash computations in SCUE. The 80\_hash and 160\_hash respectively represent the needed 80/160 cycles (normalized to the 80\_hash).}
	\label{160_writelatency}
	\vspace{-0.5cm}
\end{figure}

\begin{figure}[t]
	\centering
	\includegraphics[width=0.45\textwidth]{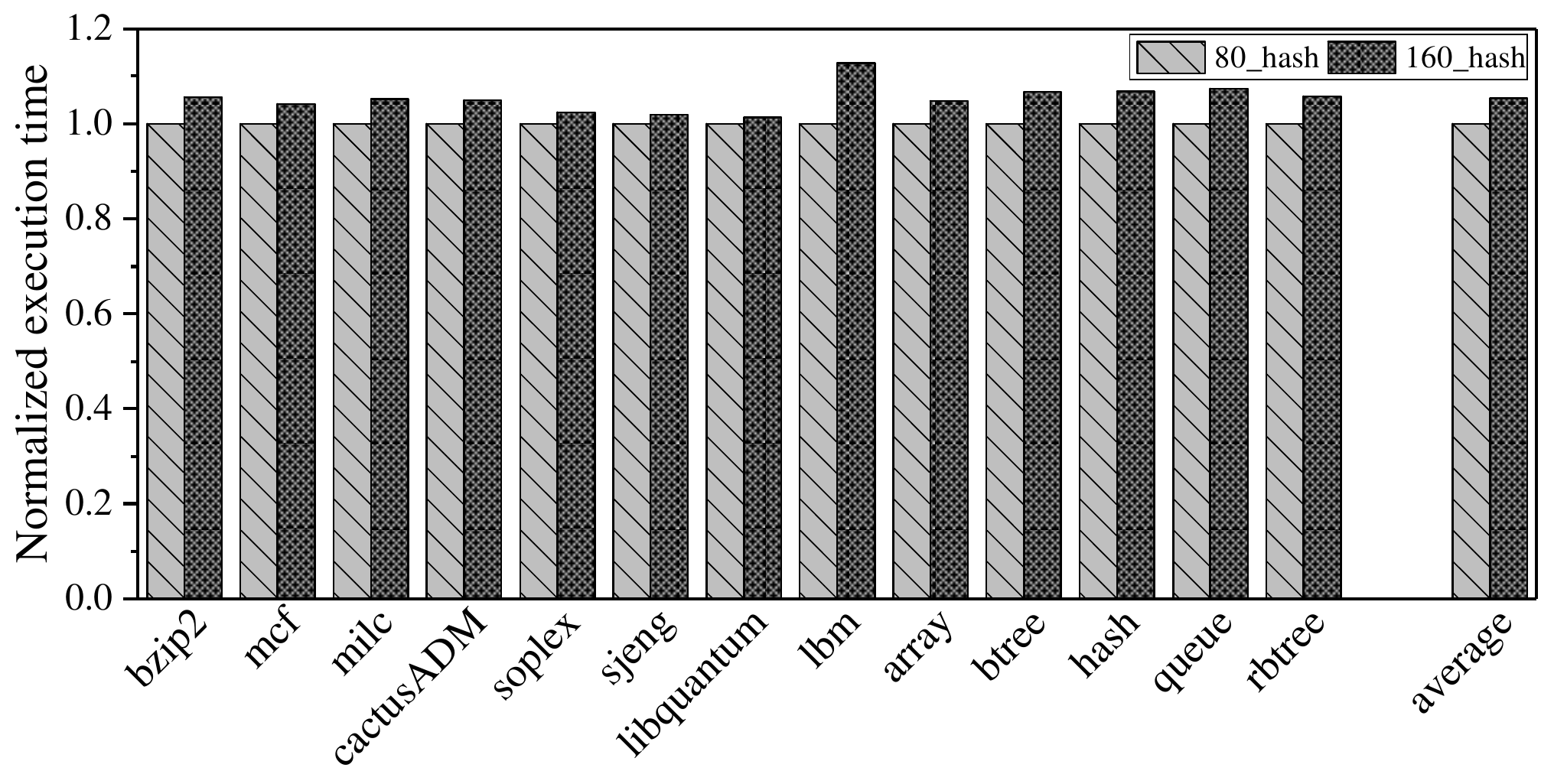}
	\vspace{-0.3cm}
	\caption{The execution time of using different hash computations in SCUE. The 80\_hash and 160\_hash respectively represent the needed 80/160 cycles (normalized to the 80\_hash).}
	\label{160_executiontime}
	\vspace{-0.3cm}
\end{figure}
\vspace{-0.1cm}
\subsection {The Sensitivity to the Hash Latency}
In our simulated configuration, the hash latency to generate the HMACs is 80 cycles. However, the hash latency can be higher in existing works~\cite{liu2019janus,suh2003efficient}, i.e., 160 cycles. We adjust the hash latency from 80 cycles to 160 cycles to evaluate the performance of SCUE.

As shown in Fig.~\ref{160_writelatency}, doubling the hash latency incurs up to 1.19x write latency in SCUE. On average, once the hash latency becomes 160 cycles, the write latency is 1.10x than that of 80-cycle hash latency configuration. Fig.~\ref{160_executiontime} shows the execution time when adjusting the hash latency. The execution time increases by 1.05x when the hash latency is 160 cycles. Since we only calculate the HMAC once when writing data, doubling the hash latency results in the slight decrease in the performance of SCUE. This result motivates us to use more secure hash algorithms with higher computation latency to protect systems~\cite{gueron2011sha} while providing high performance.

\subsection {Recovery Scheme}
\label{recoverytime}
After system failures, a system needs to recover the integrity tree for the following verification. We use Osiris~\cite{YeHA18} to recover the counter blocks. For integrity tree nodes, existing works~\cite{ZubairA19,AwadYSNZ19} offer different approaches to recovering the BMT and SIT with different overheads, since SIT can't be reconstructed from the leaf nodes. In our SCUE, since the SIT is reconstructed from the leaf nodes via our proposed counter-summing recovery approach, all existing recovery schemes, no matter whether they are designed for BMT or SIT, can be used in the shortcut updated SIT. For better recovery performance, we use the AGIT in the Anubis~\cite{ZubairA19}, which is designed for BMT, to recover the shortcut updated SIT. In fact, any recovery schemes designed for BMT and SIT can be used.

\section{Related Work}
\label{section 7}

\textbf{Security metadata recovery.} Security metadata include counter blocks and integrity tree nodes. For improving performance, security metadata are cached in a volatile on-chip buffer in the memory controller. After system failures, some updates of security metadata are lost due to not being persisted into NVM in time. To recover the counter blocks, Osiris~\cite{YeHA18} relaxes the persistence of counter blocks, and retrieves the counters from the stale state in NVM. Supermem~\cite{zuo2019supermem} uses write-through scheme to ensure the consistency of counter blocks. When the counter blocks are modified, they are directly flushed into NVM. SCA~\cite{LiuKRK18} consistently persists the counter and encrypted data via the support of ADR. To recover the BMT, Triad-NVM~\cite{AwadYSNZ19} persists low-level tree nodes with user data. After failures, the systems can reconstruct the BMT from the persistent low-level nodes. Anubis~\cite{ZubairA19} records the address and content of the modified cached metadata in the shadow table in NVM. According to the shadow table, Anubis has the ability to recover both BMT and SIT. Phoenix~\cite{alwadi2019phoenix} observed that Anubis needs 2x writes to recover SIT. To reduce the write overhead of recovering SIT, Phoenix relaxes the persistences of counter blocks via Osiris, and recovers the intermediate tree nodes via Anubis.

\textbf{Security metadata organization.} The security metadata are organized in multiple ways to improve the performance of accessing metadata. To reduce the access latency and space overhead of integrity trees, existing schemes propose multiple variants of SIT. VAULT~\cite{TaassoriSB18} reduces the height and space overhead of SIT by storing more than 8 counters in one node, e.g., 64 counters in the leaf node, 32 counters in the Level-1 node and 16 counters in Level-2 and upper-level nodes. Based on VAULT, Morphable Counters~\cite{SaileshwarNREJQ18} observed that when one counter overflows, either less than a quarter of counters or all the counters are used. The Morphable Counters scheme provides a scalable solution to store 128 counters in one node and further reduces the height of the tree. Synergy~\cite{SaileshwarNREQ18} places the MAC inside the ECC chip in a 9-chip ECC-DIMMs and demonstrates that MAC can be used to detect not only data tampering but also memory errors. ITESP~\cite{ITESP} uses small counters to save space in the tree node. The ECCs of the child nodes are XORed, and the XORed ECC is embedded in the parent node in ITESP.

\textbf{The overheads of updating the integrity tree.} Updating the integrity tree incurs large latency and consumes computation overheads. To reduce the overhead of integrity verification, Janus~\cite{liu2019janus} executes the integrity tree updates with backend operations (e.g., encryption and deduplication) in parallel. Janus also pre-executes the tree updates before the write requests arriving at the memory controller. Freij et al.~\cite{freij2020streamlining} observed that updating the BMT with correct order needs a large overhead. They propose the pipelining BMT update scheme to reduce the latency of updating BMT.

Unlike existing schemes, our SCUE directly updates the root to exhibit the changes of leaf nodes by overlooking updating the intermediate tree nodes. Therefore, SCUE delivers high performance on NVM systems while ensuring the integrity.

\section{Conclusion}
\label{section 8}
In order to correctly update the root of the integrity tree with low overheads, this paper proposes the low-latency and shortcut updated scheme SCUE. The idea behind SCUE is that only the updates in persistent leaf nodes and on-chip root are necessary. Ensuring the consistency between the root and the leaf nodes causes most overhead to persist the security metadata. Since the updates in cached intermediate tree nodes are not used, it is unnecessary to update the root in a step-by-step manner via modifying the intermediate tree nodes. We directly update the root in the SIT to significantly reduce the update overheads. To consistently update the tree root, we pre-update the root in the ADR. A counter-summing recovery approach is used to reconstruct the SIT from the consistent leaf nodes after system reboots. Compared with the eager update scheme, the SCUE shows 1.59x performance improvements in terms of system execution time while offering integrity verification after system failures and reboots.



\bibliographystyle{IEEEtranS}
\bibliography{references}

\end{document}